\documentclass[journal,onecolumn]{tran}
\pdfoutput=1
\usepackage{textcomp}
\usepackage{amsmath,amssymb,amsfonts}
\usepackage{algorithmic}
\usepackage{cite}

\usepackage[ruled]{algorithm2e}

\usepackage[american]{babel}
\usepackage{microtype}
\usepackage{graphicx}

\usepackage{mathptmx}

\usepackage{mathbbol}

\usepackage{caption}
\usepackage{framed}

\usepackage{subfigure}

\usepackage{amsmath}

\usepackage[numbers]{natbib}
\setcitestyle{open={},close={}}

\usepackage{stfloats}

\usepackage{color}

\usepackage{amsthm}

\def\BibTeX{{\rm B\kern-.05em{\sc i\kern-.025em b}\kern-.08em
    T\kern-.1667em\lower.7ex\hbox{E}\kern-.125emX}}
\begin{document}

\title{Outage Constrained Robust Secure Beamforming in Cognitive Satellite-Aerial Networks\\
}

\author{Bai~Zhao,
 Min~Lin,
 Ming~Cheng,
 Wei-Ping~Zhu,
 and ~Naofal~Al-Dhahir
\thanks{B. Zhao, M. Lin and M. Cheng  are with the College of Telecommunications and Information Engineering,
 Nanjing University of Posts and Telecommunications,
 Nanjing, China. 

W.-P. Zhu is with the Department of Electrical and Computer Engineering, Concordia University, Montreal, Canada, and also with the College of Telecommunications and Information Engineering,
Nanjing University of Posts and Telecommunications,
Nanjing, China.

N. Al-Dhahir is with the Department of Electrical and Computer Engineering, The University of Texas at Dallas, Richardson, TX 75080 USA.

Corresponding author: Min Lin (linmin@njupt.edu.cn).

This work was supported in part by the Key International Cooperation Research Project under Grant 61720106003, in part by the Shanghai Aerospace Science and Technology Innovation Foundation under Grant SAST2019-095, in part by NUPTSF under Grant NY220111, and in part by the Postgraduate Research and Practice Innovation Program of Jiangsu Province under Grant KXCY20$\_$0814.}
}

\maketitle
\vspace{-2em}
\begin{abstract}
This paper proposes a robust beamforming scheme to enhance the physical layer security (PLS) of multicast transmission in a cognitive satellite and aerial network (CSAN) operating in the millimeter wave frequency band. Based on imperfect channel state information (CSI) of both eavesdroppers (Eves) and primary users (PUs), we maximize the minimum achievable secrecy rate (ASR) of the secondary users (SUs) in the aerial network under the constraints of the interference to the PUs in the satellite network, the quality of service (QoS) requirements of the SUs and per-antenna power budget of the aerial platform. To tackle this mathematically intractable problem, we first introduce an auxiliary variable and outage constraints to simplify the complex objective function. We then convert the non-convex outage constraints into deterministic forms and adopt penalty function approach to obtain a semi-definite problem such that it can be solved in an iterative fashion. Finally, simulation results show that with the transmit power increase, the minimal ASR of SUs obtained from the proposed BF scheme well approximate the optimal value.
\end{abstract}
\vspace{-0.7em}
\vspace{-1em}
\section{Introduction}
\vspace{-0.4em}
\IEEEPARstart{S}{atellite} communication (SatCom)  with its inherent advantage of wide coverage is considered as a promising and even indispensable solution of providing seamless connection and high-speed broadband services for fixed and mobile users in remote areas [\cite {9210567-1}]-[\cite {9062335-3}]. In contrast, the high altitude platform (HAP)-based aerial communication, due to its  relatively low cost and easy deployment, is emerging as a new and attractive paradigm for the fifth generation systems as recommended by 3GPP [\cite {9062335-4}]. However, according to the recommendation of the International Telecommunication Union (ITU), the frequency range of 18-32 GHz is one of the spectrum candidates for HAP, which overlaps with the Ka band in SatCom [\cite {7060478-6}]. In this regard, the framework of cognitive satellite and aerial networks (CSAN) has been presented [\cite {8892508-7}],  and has already been adopted in the ABSOLUTE project [\cite {9314201-7}]. Nevertheless, how to efficiently eliminate the interference due to spectrum sharing is one of the urgent issues in cognitive ratio networks.
\vspace{-0.4em}

Another key issue of CSAN is the physical layer security (PLS) due to the broad coverage of HAP [\cite {7263345-9}]. In this context, beamforming (BF) technology was proposed as an effective solution to cancel the mutual interference and enhance  PLS in the considered CSAN, as it is able to align the desired user with the maximum beam gain direction and suppress signal leakage to the undesired users [9]-[13]. For instance, the authors of [\cite {8334240-10}] proposed a heuristic BF scheme to obtain the weight vector analytically by considering the scenario of coordinated eavesdropping. In [\cite {8333695-11}], a cooperative secure transmission BF scheme has been designed to solve an achievable secrecy rate (ASR) maximization problem under a total power budget constraint. The authors of [\cite {8780752-12}] obtained the optimal BF weight vector by applying semi-definite programming (SDP) and Taylor series expansion. Furthermore, the authors of [\cite {9072611-13}] proposed a novel BF scheme to degrade the possible wiretap channels efficiently. Besides, the BF schemes based on hybrid zero-forcing and partial zero-forcing were presented in [\cite {7583691-14}]. However, it should be pointed out that all of the above mentioned works focused only on the cognitive satellite and terrestrial network, while PLS in CSAN for the multicast is still an open research topic. In addition, due to the inherent complexity and uniqueness of multicast, the PLS in multicast transmission is more interesting and challenging than that of unicast.
\vspace{-0.3em}

Motivated by these observations, we propose a robust BF scheme to enhance the security in a CSAN with multicast transmission, where only imperfect channel state information (CSI) of both the eavesdroppers (Eves) and primary users (PUs) is assumed to be available. We first set up an optimization problem to maximize the minimal achievable secrecy rate of the SUs,  while satisfying the inteference power constraint on PUs in the satellite network, the quality of service (QoS) requirements of the secondary users (SUs) and the per-antenna power budget constraint at the aerial platform. To tackle this non-convex problem, we then introduce an auxiliary variable and outage constraints to simplify the complex objective function. Further, by transforming the non-convex outage constraints into convex ones and using SDP together with penalty function, we design a robust BF scheme to realize the PLS in the considered CSAN. Notably, compared with the related works in [\cite {8334240-10}]-[\cite {7583691-14}], we focus on robust BF to maximize the ASR of the SUs in secure multicast transmission, which, to the best of our knowledge, is still an open yet important problem in CSAN.

\vspace{-0.8em}
\section{System model}
\vspace{-0.4em}

In this paper, we consider the downlink transmission of a cognitive satellite and aerial network, where the HAP system, termed as a secondary network, shares the millimeter wave (mmWave) band with the geostationary orbit (GEO) satellite system, termed as a primary network. Specifically, the GEO satellite employs array fed reflector antennas with ${N_s}$ feeds to serve $Q$ PUs within the coverage region, while the HAP utilizes a uniform planar array (UPA) with  ${N_h} = {N_1} \times {N_2}$ elements and transmits signals to $M$ SUs that will be wiretapped by $K$ Eves. Here, the HAP exploits imperfect CSIs of Eves and PUs to conduct secure BF for multicast communications, which is a more general and practical scenario than the work in [\cite {8333695-11}],[\cite {8780752-12}], where perfect CSI is required to implement the BF scheme in unicast transmission. Without loss of generality, it is assumed that the interference from the GEO satellite to the SUs is negligible due to the long distance and heavy shadowing [\cite {7583691-15}].

In this paper, we consider a scenario where both the satellite and HAP transmit signals use a multicast scheme. Compared with the unicast, multicast is capable of sending packages to multiple users simultaneously with a single transmission resource, and thus satisfying the requirements of future wireless networks. In this regard, HAP delivers the intended signal $s\left( t \right)$  with unit power to SUs, which is mapped onto a uniform planar array with multicast BF weight vector ${\mathbf{w}} \in {\mathbb{C}^{{N_h} \times 1}}$. Thus, the received signal at the $m{\rm{ - th}}$ SU and $k{\rm{ - th}}$  Eve can be written as
\begin{equation}\label{eqYR}
\setlength{\abovedisplayskip}{3pt}
\setlength{\belowdisplayskip}{3pt}
{y_m}\left( t \right) = {\mathbf{h}}_m^H{\mathbf{w}}s\left( t \right) + {n_m}\left( t \right),
\end{equation}
\begin{equation}\label{eqYR}
\setlength{\abovedisplayskip}{3pt}
\setlength{\belowdisplayskip}{3pt}
{y_{e,k}}\left( t \right) = {\bf{h}}_{e,k}^H{\bf{w}}s\left( t \right) + {n_{e,k}}\left( t \right),
\end{equation}
where ${n_m}\left( t \right)$  and ${n_{e,k}}\left( t \right)$  represent independent identically distributed additive white Gaussian noises (AWGN) with variances $\sigma _m^2$  and $\sigma _{e,k}^2$, respectively. In addition,  ${{\mathbf{h}}_m} \in {\mathbb{C}^{{N_h} \times 1}}$ and  ${{\mathbf{h}}_{e,k}} \in {\mathbb{C}^{{N_h} \times 1}}$ denote the channel vectors from HAP to  $m{\rm{ - th}}$  SU and $k{\rm{ - th}}$  Eve, respectively. According to (1) and (2), the output signal-to-noise ratio (SNR) at the  $m{\rm{ - th}}$ SU and $k{\rm{ - th}}$  Eve can be expressed as
\begin{equation}\label{eqSNRR}
\setlength{\abovedisplayskip}{3pt}
\setlength{\belowdisplayskip}{3pt}
  {\gamma _m} = \frac{{{{\left| {{\bf{h}}_m^H{\bf{w}}} \right|}^2}}}{{\sigma _m^2}},
\end{equation}
\begin{equation}\label{eqSNRR}
\setlength{\abovedisplayskip}{3pt}
\setlength{\belowdisplayskip}{3pt}
{\gamma _{e,k}} = \frac{{{{\left| {{\bf{h}}_{e,k}^H{\bf{w}}} \right|}^2}}}{{\sigma _{e,k}^2}}.
\end{equation}
\vspace{-0.3em}

As a result, the achievable rate of the $m{\rm{ - th}}$  SU and $k{\rm{ - th}}$  Eve can be expressed as
\begin{equation}\label{eqSNRR}
\setlength{\abovedisplayskip}{3pt}
\setlength{\belowdisplayskip}{3pt}
{R_m} = {\log _2}\left( {1 + {\gamma _m}} \right) = {\log _2}\left( {1 + \frac{{{{\left| {{\bf{h}}_m^H{\bf{w}}} \right|}^2}}}{{\sigma _m^2}}} \right),
\end{equation}
\begin{equation}\label{eqSNRR}
\setlength{\abovedisplayskip}{3pt}
\setlength{\belowdisplayskip}{3pt}
{R_{e,k}} = {\log _2}\left( {1 + {\gamma _{e,k}}} \right) = {\log _2}\left( {1 + \frac{{{{\left| {{\bf{h}}_{e,k}^H{\bf{w}}} \right|}^2}}}{{\sigma _{e,k}^2}}} \right).
\end{equation}
Consequently, the achievable secrecy rate of the  $m{\rm{ - th}}$ SU is given by
\begin{equation}\label{eqSNRR}
\setlength{\abovedisplayskip}{3pt}
\setlength{\belowdisplayskip}{3pt}
{R_{s,m}} = {\left[ {{{\log }_2}\left( {1 + {\gamma _m}} \right) - \mathop {\max }\limits_k {{\log }_2}\left( {1 + {\gamma _{e,k}}} \right)} \right]^ + },
\end{equation}
where ${\left[ x \right]^ + } = \max \left( {x,0} \right)$. Here, we are only interested in the practical scenario where the ASR is nonnegative. Hence, the superscript “+” can be removed.

Considering the highly directional characteristic of mmWave communications, the HAP downlink channel can be modeled as the superposition of a predominant line-of-sight (LoS) component and a sparse set of single-bounce non-LoS (NLoS) components. To this end, the channel model of the HAP network can be written as [\cite {8334240-10}]
\begin{equation}\label{eqYk}
\setlength{\abovedisplayskip}{3pt}
\setlength{\belowdisplayskip}{3pt}
\begin{array}{l}
{\bf{h}} = \sqrt {g\left( {\theta ,\varphi } \right)} {\rho _0}{{\bf{a}}_a}\left( {\theta ,\varphi } \right) \otimes {{\bf{a}}_e}\left( {\theta ,\varphi } \right)\\
\;\;\;\;\;\;\; + \sqrt {\frac{1}{L}} \sum\limits_{l = 1}^L {\sqrt {g\left( {{\theta _l},{\varphi _l}} \right)} {\rho _l}} {{\bf{a}}_a}\left( {{\theta _l},{\varphi _l}} \right) \otimes {{\bf{a}}_e}\left( {{\theta _l},{\varphi _l}} \right),
\end{array}
\end{equation}
where ${\rho _0}$ and ${\rho _l}$ denote, respectively, the propagation losses for the LoS component and the $l{\rm{ - th}}$  NLoS components. Typically, the LoS path loss is 5 dB to 10 dB larger than the NLoS path losses. And $L$ denotes the number of NLoS paths. In addition, $g\left( {\theta ,\varphi } \right)$  denotes the directivity pattern with $\theta $  being the pitching angle and $\varphi $ the azimuthal angle. According to the model introduced by the ITU [\cite {19}], the directivity pattern in dB, namely,  ${\rm
{\hat g}}\left( {\theta ,\varphi } \right) = 10{\log _{10}}{\text{g}}\left( {\theta ,\varphi } \right)$, is computed by
\begin{equation}\label{eqSNRDk}
\setlength{\abovedisplayskip}{3pt}
\setlength{\belowdisplayskip}{3pt}
{\rm{\hat g}}\left( {\theta ,\varphi } \right) = {G_{\max }} - \min \left\{ {{{\rm{g}}_a}\left( {\theta ,\varphi } \right) + {{\rm{g}}_e}\left( {\theta ,\varphi } \right),\;SLL} \right\},
\end{equation}
where ${G_{\max }}$  represents the maximum antenna gain, and $SLL$ denotes the side-lobe level of the antenna pattern. Moreover, ${{\text{g}}_a}\left( {\theta ,\varphi } \right)$  and ${{\text{g}}_e}\left( {\theta ,\varphi } \right)$  are the relative patterns along the X-axis and Y-axis, respectively, and can be expressed as [\cite {8334240-10}]
\begin{equation}\label{eqphi}
\setlength{\abovedisplayskip}{3pt}
\setlength{\belowdisplayskip}{3pt}
{g_a}\left( {\theta ,\varphi } \right) = \min \left\{ {12{{\left( {\frac{{\arctan \left( {{{\cot \theta } \mathord{\left/
 {\vphantom {{\cot \theta } {\cos \varphi }}} \right.
 \kern-\nulldelimiterspace} {\cos \varphi }}} \right)}}{{\varphi _a^{3{\rm{dB}}}}}} \right)}^2},SLL} \right\},
\end{equation}
\begin{equation}\label{eqSNRk}
\setlength{\abovedisplayskip}{3pt}
\setlength{\belowdisplayskip}{3pt}
{g_e}\left( {\theta ,\varphi } \right) = \min \left\{ {12{{\left( {\frac{{\arctan \left( {\tan \theta \sin \varphi } \right)}}{{\varphi _e^{3{\rm{dB}}}}}} \right)}^2},SLL} \right\},
\end{equation}
where $\varphi _a^{3{\rm{dB}}}$  and  $\varphi _e^{3{\rm{dB}}}$ represent the 3 dB beamwidths of the X-axis and Y-axis, respectively. In (8),  ${{\bf{a}}_a}\left( {\theta ,\varphi } \right)$ and  ${{\bf{a}}_e}\left( {\theta ,\varphi } \right)$ are array steering vectors associated with the X-axis and Y-axis, respectively, and can be expressed as
\begin{equation}\label{eqSNRk}
\setlength{\abovedisplayskip}{3pt}
\setlength{\belowdisplayskip}{3pt}
{{\bf{a}}_a}\left( {\theta ,\varphi } \right) = {\left[ {1,{e^{jk{d_1}\sin \theta \cos \varphi }},...,{e^{jk{d_1}\left( {{N_1} - 1} \right)\sin \theta \cos \varphi }}} \right]^T},
\end{equation}
\begin{equation}\label{eqSNRk}
\setlength{\abovedisplayskip}{3pt}
\setlength{\belowdisplayskip}{3pt}
{{\bf{a}}_e}\left( {\theta ,\varphi } \right) = {\left[ {1,{e^{jk{d_2}\sin \theta \sin \varphi }},...,{e^{jk{d_2}\left( {{N_2} - 1} \right)\sin \theta \sin \varphi }}} \right]^T},
\end{equation}
where  $k = {{2\pi } \mathord{\left/
 {\vphantom {{2\pi } \lambda }} \right.
 \kern-\nulldelimiterspace} \lambda }$ is the wavenumber,  $\lambda $ being the wavelength,  and ${d_1}$ and ${d_2}$  are the antenna element spacings along the X-axis and that of Y-axis, respectively. In the next section, we will propose a robust BF scheme to realize PLS in the considered CSAN.

\vspace{-1em}
\section{Proposed robust secure bf scheme}
\vspace{-0.3em}
In the existing works [\cite {8333695-11}], [\cite {8780752-12}], BF schemes are designed based on perfect CSI of both Eves and PU, which is impractical because of channel estimation error, feedback quantization error, HAP disturbance. Thus, here we consider a more practical case where only imperfect CSI is available at the HAP and the corresponding channel vectors of Eves and PUs\footnote{In this paper, we assume that the Eves are inner-system users but untrusted by legitimate receivers, such that the HAP can estimate the CSI of the Eves by training and feedback [\cite {6819814-16-R}], just like legitimate users.}, namely ${{\bf{h}}_{e,k}}$  and ${{\bf{h}}_q}$, can be modeled as
\begin{equation}\label{eqhSR}
\setlength{\abovedisplayskip}{3pt}
\setlength{\belowdisplayskip}{3pt}
{{\bf{h}}_{e,k}} = {{\bf{\hat h}}_{e,k}} + \Delta {{\bf{\tilde h}}_{e,k}},
\end{equation}
\begin{equation}\label{eqhSR}
\setlength{\abovedisplayskip}{3pt}
\setlength{\belowdisplayskip}{3pt}
  {{\bf{h}}_q} = {{\bf{\hat h}}_q} + \Delta {{\bf{\tilde h}}_q},
\end{equation}
where ${{\bf{\hat h}}_{e,k}}$  and  ${{\bf{\hat h}}_q}$ are the estimated channel vectors associated with ${\rm{Ev}}{{\rm{e}}_k}$  and ${\rm{P}}{{\rm{U}}_q}$, respectively. Herein,  $\Delta {{\bf{\tilde h}}_{e,k}}$ and $\Delta {{\bf{\tilde h}}_q}$ are the corresponding stochastic channel vectors measuring the channel estimation errors, and follow the complex Gaussian distribution, i.e.,  $\Delta {{\mathbf{\tilde h}}_{e,k}} \sim {\cal C}{\cal N}\left( {0,{{\mathbf{E}}_{e,k}}} \right)$
, ${{\bf{E}}_{e,k}} \succ 0$ and $\Delta {{\mathbf{\tilde h}}_q} \sim {\cal C}{\cal N}\left( {0,{{\mathbf{E}}_q}} \right)$, ${{\bf{E}}_q} \succ 0$, where  ${\cal C}{\cal N}\left( {0,{\pmb{\Sigma }}} \right)$ with mean zero and covariance matrix $\pmb{\Sigma }$.

To guarantee the secure transmission of each SU, we conduct a robust multicast BF design by maximizing the worst-case ASR, while satisfying the minimum SNR constraint for SUs and outage probability constraint of interference power for PUs. Since the latter is caused by the leakage of energy during the communication between HAP and SUs, the interference power can be calculated using  ${\left| {{\mathbf{h}}_q^H{\mathbf{w}}} \right|^2}$. In addition, the per-antenna power constraint of the HAP is used in our work, which is more practical compared to the total power constraint due to the lack of flexibility of transmit power sharing among elements. To this end, the constrained optimization problem can be formulated as
\begin{equation}\label{eqhSR}
\setlength{\abovedisplayskip}{3pt}
\setlength{\belowdisplayskip}{3pt}
\begin{array}{l}
\mathop {\max }\limits_{\bf{w}} \;\mathop {\min }\limits_m \;\mathop {\min }\limits_k \;\mathop {\min }\limits_{\Delta {{{\bf{\tilde h}}}_{e,k}},\Delta {{{\bf{\tilde h
}}}_q}} \left\{ {{R_m} - {R_{e,k}}} \right\}\\
{\rm{s}}{\rm{.t}}{\rm{.}}\;\;\;\;\;\Pr \left\{ {{{\left| {{\bf{h}}_q^H{\bf{w}}} \right|}^2} \le {{\rm{I}}_{th}}} \right\} \ge 1 - {{\rm{P}}_{out,1}},\forall q,\\
\;\;\;\;\;\;\;\;\;\;\;{\gamma _m} \ge {\gamma _{th}},\;\forall m,\\
\;\;\;\;\;\;\;\;\;\;\;{\left| {{{\left[ {\bf{w}} \right]}_n}} \right|^2} \le {P_n},\;\forall n,
\end{array}
\end{equation}
where  ${{\rm{I}}_{th}}$ and ${\gamma _{th}}$  denote the interference power threshold for primary users and the SNR threshold for all SUs, respectively,  ${P_{out,1}}$ denotes the outage probability threshold, and  ${P_n}$ denotes the power budget of the $n{\rm{ - th}}$  antenna of the HAP. 

Obviously, the original optimization problem (16) is difficult to solve directly due to its complex objective function and non-convex outage constraint. To simplify this objective function, we first introduce an auxiliary variable $R$  and define ${\bf{W}} = {\bf{w}}{{\bf{w}}^H}$. Then, the constrainted optimization problem can be reformulated as
\begin{subequations}\label{eqhSR}
\setlength{\abovedisplayskip}{3pt}
\setlength{\belowdisplayskip}{4pt}
\begin{align}
\mathop {\max }\limits_{{\mathbf{W}}\underset{\raise0.3em\hbox{$\smash{\scriptscriptstyle-}$}}{ \succ } 0,R} \;\;\;\;\;\;\;\;\;\;\;\;\;\;\;\;\;\;\;R\;\;\;\;\;\;\;\;\;\;\;\;\;\;\;\;\;\;\;\;\;\;\;\;\;\;\;\;\;\;\; \;\;\;\;\;\;\;\;\;\;\;\;\;\;\;\;\;\;\;\;\;\;\;\;\;\;\;\;\;\;\;\;\;\;\; \\
{\rm{s}}{\rm{.t}}{\rm{.}}\;\;\;\;\Pr \left\{ {{{ {{\bf{h}}_q^H{\bf{W}}{{\bf{h}}_q}} }} \le {{\rm{I}}_{th}}} \right\} \ge 1 - {{\rm{P}}_{out,1}},\;\forall q,\;\;\;\;\;\;\;\;\;\;\;\;\;\;\;\;\;  \\
\Pr \left\{ {{{\log }_2}\left( {\frac{{1 + \frac{{{\bf{h}}_m^H{\bf{W}}{{\bf{h}}_m}}}{{\sigma _m^2}}}}{{1 + \frac{{{\bf{h}}_{e,k}^H{\bf{W}}{{\bf{h}}_{e,k}}}}{{\sigma _{e,k}^2}}}}} \right) \ge R} \right\} \ge 1 - {{\rm{P}}_{out,2}},\;\forall m,k, \\
\;\frac{{{\bf{h}}_m^H{\bf{W}}{{\bf{h}}_m}}}{{\sigma _m^2}} \ge {\gamma _{th}},\;\forall m,\;\;\;\;\;\;\;\;\;\;\;\;\;\;\;\;\;\;\;\;\;\;\;\;\;\;\;\;\;\;\;\;\;\;\;\;\;\;\;  \\
\left| {{{\left[ {\bf{W}} \right]}_{nn}}} \right|\; \le {P_n},\;\forall n,\;\;\;\;\;\;\;\;\;\;\;\;\;\;\;\;\;\;\;\;\;\;\;\;\;\;\;\;\;\;\;\;\;\;\;\;\;\;\;\;\;\;\; \\
\;\;{\rm{rank}}\left( {\bf{W}} \right) = 1.\;\;\;\;\;\;\;\;\;\;\;\;\;\;\;\;\;\;\;\;\;\;\;\;\;\;\;\;\;\;\;\;\;\;\;\;\;\;\;\;\;\;\;\;\;\;\;\;
\end{align}
\end{subequations}

\vspace{-0.5em}
Note that the above problem is a one-dimensional search problem associated with $R$, which is equivalent to determining the feasibility of ${\bf{W}}$  that satisfies
\begin{subequations}\label{eqhSR}
\setlength{\abovedisplayskip}{4pt}
\setlength{\belowdisplayskip}{4pt}
\begin{align}
&{\text{find}}\;\;\;\;\; {\mathbf{W}}\underset{\raise0.3em\hbox{$\smash{\scriptscriptstyle-}$}}{ \succ } 0 \\
&{\rm{s}}{\rm{.t}}{\rm{.}} \left( {{\rm{17b}}} \right) \sim \left( {1{\rm{7f}}} \right). 
\end{align}
\end{subequations}

\vspace{-0.5em}
In the following, we will transform several non-convex constraints, i.e. (17b), (17c) and (17f), into convex ones. Firstly, we focus on the outage constraint (17c). By rewriting the CSI error as $\Delta {{\bf{h}}_{e,k}} = {\bf{E}}_{h,k}^{{1 \mathord{\left/
 {\vphantom {1 2}} \right.
 \kern-\nulldelimiterspace} 2}}{{\bf{v}}_{h,k}}$, where ${{\mathbf{v}}_{h,k}}\! \sim \! \;{\cal C} {\cal N}\left( {0,{{\mathbf{I}}_{{N_h}}}} \right)$ and  ${{\mathbf{I}}_{{N_h}}}$ denotes the ${N_h}$-dimensions identity matrix, the constraint (17c) can be expressed as
\begin{equation}\label{eqhSR}
\setlength{\abovedisplayskip}{4pt}
\setlength{\belowdisplayskip}{4pt}
\begin{aligned}
\begin{array}{l}
\Pr \left\{ {{\bf{v}}_{h,k}^H{{\bf{Q}}_{h,k}}{{\bf{v}}_{h,k}} + 2{\mathop{\rm Re}\nolimits} \left\{ {{\bf{v}}_{h,k}^H{{\bf{b}}_{h,k}}} \right\} \ge } \right.\left. {\frac{{\sigma _{e,k}^2}}{{\sigma _m^2}}{2^{ - R}}\left( {{\bf{h}}_m^H{\bf{W}}{{\bf{h}}_m} + \sigma _m^2} \right) - {\bf{\hat h}}_{e,k}^H{\bf{W}}{{{\bf{\hat h}}}_{e,k}} - \sigma _{e,k}^2} \right\} \le {{\rm{P}}_{out,2}},
\end{array}
\end{aligned}
\end{equation}
where ${{\bf{Q}}_{h,k}} = {\left( {{\bf{E}}_{h,k}^{{1 \mathord{\left/
 {\vphantom {1 2}} \right.
 \kern-\nulldelimiterspace} 2}}} \right)^H}{\bf{WE}}_{h,k}^{{1 \mathord{\left/
 {\vphantom {1 2}} \right.
 \kern-\nulldelimiterspace} 2}}$  and ${{\bf{b}}_{h,k}} = {\bf{E}}_{h,k}^{{1 \mathord{\left/
 {\vphantom {1 2}} \right.
 \kern-\nulldelimiterspace} 2}}{\bf{W}}{{\bf{\hat h}}_{h,k}}$.

\vspace{-0.4em}
Since the above outage constraint is still non-convex, we employ the Bernstein-Type Inequality I to transform (19) into a convex one [\cite {6891348-16}]. The Bernstein-Type Inequality I has the capability to bound the tail probability of quadratic forms of Gaussian variables involving matrices, so as to convert the outage constraint into a deterministic form. By introducing an auxiliary variable $\tau $, the outage constraint (19) can be reformulated as the following deterministic form
\begin{subequations}\label{eqhSR}
\setlength{\abovedisplayskip}{4pt}
\setlength{\belowdisplayskip}{4pt}
\begin{align}
\begin{gathered}
{\delta _{h,k}} \geq {\text{Tr}}\left( {{{\mathbf{Q}}_{h,k}}} \right) + \sqrt {2{\sigma _h}} \sqrt {{{\left\| {{\text{vec}}\left( {{{\mathbf{Q}}_{h,k}}} \right)} \right\|}^2} + 2{{\left\| {{{\mathbf{b}}_{h,k}}} \right\|}^2}}  \hfill \\
  \;\;\;\;\;\;\;\;\;\;\;\; + {\sigma _h}{u^ + }\left( {{{\mathbf{Q}}_{h,k}}} \right),\;\forall k, \hfill \\ 
\end{gathered} \\
\tau  \leq {\mathbf{h}}_m^H{\mathbf{W}}{{\mathbf{h}}_m} + \sigma _m^2,\;\forall m, \;\;\;\;\;\;\;\;\;\;\;\;\;\;\;\;\;\;\;\;\;\;\;\;\;
\end{align}
\end{subequations}
where  ${\delta _{h,k}} = \frac{{\sigma _{e,k}^2}}{{\sigma _m^2}}{2^{ - R}}\tau  - {\bf{\hat h}}_{e,k}^H{\bf{W}}{{\bf{\hat h}}_{e,k}} - \sigma _{e,k}^2,\;\forall k$ and ${\sigma _h} =  - \ln \left( {{P_{out,2}}} \right)$. ${u^ + }\left( {\bf{A}} \right) = \max \left[ {{\lambda _{\max }}\left( {\bf{A}} \right),0} \right]$ with ${{\lambda _{\max }}\left( {\bf{A}} \right)}$ denotes the maximum eigenvalue of maxtrix ${\bf{A}}$. By introducing some auxiliary variables, i.e. ${\alpha _k}$ and ${\xi _{h,k}}$,  into (20a), we have
\begin{subequations}\label{eqhSR}
\setlength{\abovedisplayskip}{4pt}
\setlength{\belowdisplayskip}{4pt}
\begin{align}
{\rm{Tr}}\left( {{{\bf{Q}}_{h,k}}} \right) + \sqrt {2{\sigma _h}} {\alpha _k} + {\sigma _h}{\xi _{h,k}} - {\delta _{h,k}} \le 0,\;\forall k, \\
\;\left\| {\begin{array}{*{20}{c}}
{vec\left( {{{\bf{Q}}_{h,k}}} \right)}\\
{\sqrt {\rm{2}} {{\bf{b}}_{h,k}}}
\end{array}} \right\| \le {\alpha _k},\;\forall k, \;\;\;\;\;\;\;\;\;\;\;\;\;\;\; \\
{\xi _{h,k}}{{\mathbf{I}}_{{N_h}}} - {{\mathbf{Q}}_{h,k}}\underset{\raise0.3em\hbox{$\smash{\scriptscriptstyle-}$}}{ \succ } 0,\;\;{\xi _{h,k}} \ge {\rm{0}}, \;\;\forall k.\;\;\;\;\;\;\;\;\;\;\;\;\;
\end{align}
\end{subequations}
In a similar manner, the outage constraint (17b) can be reformulated as
\begin{subequations}\label{eqhSR}
\setlength{\abovedisplayskip}{4pt}
\setlength{\belowdisplayskip}{4pt}
\begin{align}
{\rm{Tr}}\left( {{{\bf{Q}}_{g,q}}} \right) + \sqrt {2{\sigma _g}} {\beta _q} + {\sigma _g}{\xi _{g,q}} - {\delta _{g,q}} \le 0,\;\forall q, \\
\left\| {\begin{array}{*{20}{c}}
{vec\left( {{{\bf{Q}}_{g,q}}} \right)}\\
{\sqrt {\rm{2}} {{\bf{b}}_{g,q}}}
\end{array}} \right\| \le {\beta _q},\;\forall q, \;\;\;\;\;\;\;\;\;\;\;\;\;\;\;\\
{\xi _{g,q}}{{\mathbf{I}}_{{N_h}}} - {{\mathbf{Q}}_{g,q}}\underset{\raise0.3em\hbox{$\smash{\scriptscriptstyle-}$}}{ \succ } 0,\;\;{\xi _{g,q}} \ge 0,\;\forall q, \;\;\;\;\;\;\;\;\;\;\;\;\;
\end{align}
\end{subequations}
where ${{\bf{Q}}_{g,q}} = {\left( {{\bf{E}}_{g,q}^{{1 \mathord{\left/
 {\vphantom {1 2}} \right.
 \kern-\nulldelimiterspace} 2}}} \right)^H}{\bf{WE}}_{g,q}^{{1 \mathord{\left/
 {\vphantom {1 2}} \right.
 \kern-\nulldelimiterspace} 2}}$,  ${{\bf{b}}_{g,q}} = {\bf{E}}_{g,q}^{{1 \mathord{\left/
 {\vphantom {1 2}} \right.
 \kern-\nulldelimiterspace} 2}}{\bf{W}}{{\bf{\hat h}}_q}$,  ${\delta _{g,q}} = {{\rm{I}}_{th}} - {\bf{\hat h}}_q^H{\bf{W}}{{\bf{\hat h}}_q}$ and  ${\sigma _g} =  - \ln \left( {{P_{out,1}}} \right)$. Thus, the one-dimensional search problem (18) is rewritten as
\begin{equation}\label{eqhSR}
\setlength{\abovedisplayskip}{4pt}
\setlength{\belowdisplayskip}{10pt}
\begin{gathered}
  {\text{find}}\;\;\;\;\;\;\;\;\;\;\;\;\;{\mathbf{W}}\underset{\raise0.3em\hbox{$\smash{\scriptscriptstyle-}$}}{ \succ } 0 \hfill \\
  {\text{s}}{\text{.t}}{\text{.}}\;\;\left( {{\text{17d}}} \right) \sim \left( {{\text{17f}}} \right){\text{,}}\left( {{\text{20b}}} \right){\text{,}}\left( {{\text{21a}}} \right) \sim \left( {{\text{21c}}} \right){\text{,}}\left( {{\text{22a}}} \right) \sim \left( {{\text{22c}}} \right). \hfill \\ 
\end{gathered}
\end{equation}

\vspace{-0.8em}
We now deal with constraint (17f) which is the last hurdle  in problem (23) to circumvent. It should be mentioned that in traditional semi-definite relaxation (SDR) approach, the rank-1 constraint is often dropped to simplify the original optimization problem. Then, the best solution will be chosen from a large number of rank-1 weight vectors that are generated randomly. Obviously, this approximate optimal solution could be suboptimal or even far from the optimal one, since the candidates from the randomization approach cannot be guaranteed to include an optimal solution of the original semi-definite problem [\cite {5447068-17}]. Here, we apply a nonsmooth method and rewrite the constraint (17f) as
\begin{equation}\label{eqhSR}
\setlength{\abovedisplayskip}{3pt}
\setlength{\belowdisplayskip}{3pt}
{\rm{Tr}}\left( {\bf{W}} \right) - {\lambda _{\max }}\left( {\bf{W}} \right) \le 0.
\end{equation}

Note that ${\rm{Tr}}\left( {\bf{W}} \right) - {\lambda _{\max }}\left( {\bf{W}} \right) \ge 0$  holds true for any semi-definite matrix ${\mathbf{W}}\underset{\raise0.3em\hbox{$\smash{\scriptscriptstyle-}$}}{ \succ } 0$, and thus (24) is equivalent to  ${\rm{Tr}}\left( {\bf{W}} \right) = {\lambda _{\max }}\left( {\bf{W}} \right)$. By adopting the penalty function approach [\cite {8892508-7}], the problem (23) is equivalent to
\begin{equation}\label{eqhSR}
\setlength{\abovedisplayskip}{3pt}
\setlength{\belowdisplayskip}{3pt}
\begin{array}{l}
\mathop {\max }\limits_{{\mathbf{W}}\underset{\raise0.3em\hbox{$\smash{\scriptscriptstyle-}$}}{ \succ } 0,\tau ,{\alpha _k},{\beta _q},{\xi _{h,k}},{\xi _{g,q}}} \; - \eta \left[ {{\rm{Tr}}\left( {\bf{W}} \right) - {\lambda _{\max }}\left( {\bf{W}} \right)} \right]\\
\;\;\;\;\;\;{\rm{s}}{\rm{.t}}{\rm{.}}\;\left( {{\rm{17d}}} \right) \sim \left( {{\rm{17e}}} \right){\rm{,}}\left( {{\rm{20b}}} \right){\rm{,}}\left( {{\rm{21a}}} \right) \sim \left( {{\rm{21c}}} \right){\text{,}}\left( {{\rm{22a}}} \right) \sim \left( {{\rm{22c}}} \right).
\end{array}
\end{equation}
where $\eta  > 0$  denotes the weight of the penalty function. Note that if we want to obtain the optimal solution of problem (25),   the value of  $\eta $ should be large enough to guarantee ${\rm{Tr}}\left( {\bf{W}} \right) - {\lambda _{\max }}\left( {\bf{W}} \right) \approx 0$. Considering that ${\lambda _{\max }}\left( {\bf{W}} \right)$  is a nonsmooth function, we employ the sub-gradient version of the maximal eigenvalue function expressed as $\partial {\lambda _{\max }}\left( {\bf{X}} \right) = {{\bf{x}}_{\max }}{\bf{x}}_{\max }^H$  and obtain
\begin{equation}\label{eqhSR}
\setlength{\abovedisplayskip}{3pt}
\setlength{\belowdisplayskip}{3pt}
{\lambda _{\max }}\left( {\mathbf{X}} \right) - {\lambda _{\max }}\left( {\mathbf{W}} \right) \geq \left\langle {{{\mathbf{w}}_{\max }}{\mathbf{w}}_{\max }^H,{\mathbf{X}} - {\mathbf{W}}} \right\rangle ,\;\forall {\mathbf{X}}\underset{\raise0.3em\hbox{$\smash{\scriptscriptstyle-}$}}{ \succ } 0,
\end{equation}
where $\left\langle {{\bf{A}},{\bf{B}}} \right\rangle  = {\rm{Tr}}\left( {{\bf{A}}{{\bf{B}}^H}} \right)$. Therefore, by defining  ${{\bf{W}}^{\left( j \right)}}$ as the optimal solution of problem (25) and calculating its maximal eigenvalue and the corresponding unit eigenvector, we have the following semi-definite problem
\begin{equation}\label{eqhSR}
\begin{aligned}
\begin{gathered}
  \mathop {\max }\limits_{{\mathbf{W}}\underset{\raise0.3em\hbox{$\smash{\scriptscriptstyle-}$}}{ \succ } 0,\tau ,{\alpha _k},{\beta _q},{\xi _{h,k}},{\xi _{g,q}}} \; - \eta \left[ {{\text{Tr}}\left( {\mathbf{W}} \right) - {\lambda _{\max }}\left( {{{\mathbf{W}}^{\left( j \right)}}} \right) - } \right. \hfill \\
  \;\;\;\;\;\;\;\;\;\;\;\;\;\;\;\;\;\;\;\;\;\;\;\;\;\;\;\;\;\;\;\; \left. {\left\langle {{\mathbf{w}}_{\max }^{\left( j \right)}{\mathbf{w}}_{\max }^{\left( j \right)H},{\mathbf{W}} - {{\mathbf{W}}^{\left( j \right)}}} \right\rangle } \right] \hfill \\
  \;\;\;\;\;\;{\text{s}}{\text{.t}}{\text{.}}\;\;\left( {{\text{17d}}} \right) \sim \left( {{\text{17e}}} \right){\text{,}}\left( {{\text{20b}}} \right){\text{,}}\left( {{\text{21a}}} \right) \sim \left( {{\text{21c}}} \right){\text{,}}\left( {{\text{22a}}} \right) \sim \left( {{\text{22c}}} \right). \hfill \\ 
\end{gathered} 
\end{aligned}
\end{equation}
Assuming that ${{\bf{W}}^{\left( {j + 1} \right)}}$  is the optimal solution of problem (25) at the  $j{\rm{ - th}}$ iteration and defining $F\left( {{{\bf{W}}^{\left( {j + 1} \right)}}} \right) =  - \eta \left[ {{\rm{Tr}}\left( {{{\bf{W}}^{\left( {j + 1} \right)}}} \right) - {\lambda _{\max }}\left( {{{\bf{W}}^{\left( {j + 1} \right)}}} \right)} \right]$ , we have
\begin{equation}\label{eqhSR}
\begin{array}{l}
F\left( {{{\bf{W}}^{\left( {j + 1} \right)}}} \right) =  - \eta \left[ {{\rm{Tr}}\left( {{{\bf{W}}^{\left( {j + 1} \right)}}} \right) - {\lambda _{\max }}\left( {{{\bf{W}}^{\left( {j + 1} \right)}}} \right)} \right]\\
\;\;\;\;\;\;\;\;\;\;\;\;\;\; \ge  - \eta \left[ {{\rm{Tr}}\left( {{{\bf{W}}^{\left( {j + 1} \right)}}} \right) - {\lambda _{\max }}\left( {{{\bf{W}}^{\left( j \right)}}} \right) - \left\langle {{\bf{w}}_{\max }^{\left( j \right)}{\bf{w}}_{\max }^{\left( j \right)H},{{\bf{W}}^{\left( {j + 1} \right)}} - {{\bf{W}}^{\left( j \right)}}} \right\rangle } \right]\\
\;\;\;\;\;\;\;\;\;\;\;\;\;\; \ge  - \eta \left[ {{\rm{Tr}}\left( {{{\bf{W}}^{\left( j \right)}}} \right) - {\lambda _{\max }}\left( {{{\bf{W}}^{\left( j \right)}}} \right)} \right]\\
\;\;\;\;\;\;\;\;\;\;\;\;\;\; = F\left( {{{\bf{W}}^{\left( j \right)}}} \right).
\end{array}
\end{equation}
\begin{algorithm}[h]
\caption{ Proposed Robust Secure BF Scheme for Multicast Transmission}
\LinesNumbered 
\KwIn{$\left\{ {{{{\mathbf{\hat h}}}_m},{{{\mathbf{\hat h}}}_{e,k}},{{{\mathbf{\hat h}}}_q},{\gamma _{th}},{{\text{I}}_{th}},{P_{out,1}},{P_{out,2}}} \right\}$}
Set tolerance of accuracy  ${\varepsilon _{\text{1}}} > 0$, and rate bounds ${R_L}$  and ${R_U}$  such that  ${R_{opt}} \in \left[ {{R_L},{R_U}} \right]$\;
Set weight  $\eta  > 0$, the convergence tolerances  ${\varepsilon _2} > 0$\;
\Repeat{${R_U} - {R_L} \leq {\varepsilon _1}$}{
Let  $R = {{\left( {{R_L} + {R_U}} \right)} \mathord{\left/
 {\vphantom {{\left( {{R_L} + {R_U}} \right)} 2}} \right.
 \kern-\nulldelimiterspace} 2}$; initialize  $j = 0$\;
Calculate ${{\mathbf{W}}^{\left( 0 \right)}}$  satisfying  $\left( {{\text{17d}}} \right) \sim \left( {{\text{17e}}} \right){\text{,}}\left( {{\text{20b}}} \right){\text{,}}\left( {{\text{21a}}} \right) \sim \left( {{\text{21c}}} \right){\text{,}}\left( {{\text{22a}}} \right) \sim \left( {{\text{22c}}} \right)$, and its maximal eigenvalue ${\lambda _{\max }}\left( {{{\mathbf{W}}^{\left( 0 \right)}}} \right)$  and corresponding eigenvector  ${\mathbf{w}}_{\max }^{\left( 0 \right)}$\;
\Repeat{$\left| {{\text{Tr}}\left( {{{\mathbf{W}}^{\left( j \right)}}} \right) - {\lambda _{\max }}\left( {{{\mathbf{W}}^{\left( j \right)}}} \right)} \right| \leq {\varepsilon _{\text{2}}}$}{
Solve (27) to obtain  ${{\mathbf{W}}^{\left( {j + 1} \right)}}$ through standard software package\;
\eIf  {${{\mathbf{W}}^{\left( {j + 1} \right)}} \approx {{\mathbf{W}}^{\left( j \right)}}$}{
Set $\eta : = 2\eta $;}{
Set $j: = j + 1$\;
Calculate ${{\mathbf{W}}^{\left( {j + 1} \right)}}$, ${\lambda _{\max }}\left( {{{\mathbf{W}}^{\left( {j + 1} \right)}}} \right)$  and the corresponding eigenvector ${\mathbf{w}}_{\max }^{\left( {j + 1} \right)}$;
}
}
\eIf  {$\;\left| {{{\left[ {\mathbf{W}} \right]}_{nn}}} \right| \leq {P_n},\;\forall n$}{
${R_L} = R$;}{
${R_U} = R$;
}
}
Calculate ${{\mathbf{w}}_{opt}}$  by using singular value decomposition (SVD) to  ${{\mathbf{W}}^{\left( j \right)}}$\;
\KwOut{robust BF weight vector ${{\mathbf{w}}} = {{\mathbf{w}}_{opt}}$. }
\end{algorithm}

\vspace{-1em}
From the above discussion, the proposed iterative procedure is convergent. Consequently, (27) can be iteratively solved with the aid of a standard convex optimization software package such as CVX, and the proposed robust BF scheme is summarized as Algorithm 1. In practice, the optimal ASR, i.e. ${R_{opt}}$,  is nonnegative, so we can set ${R_L}$ as zero in  Algorithm 1. On the other hand, we can set ${R_U}$ as a predetermined number which is larger enough to make  ${R_{opt}}$ located in the interval of $\left[ {{R_L},{R_U}} \right]$.

According to [\cite {6891348-16}], the computational complexity per-iteration is mainly caused by the number of optimization variables, the number of linear matrix inequality (LMI) constraints and their size, the number of the second-order cone (SOC) constraints and their size. For the proposed robust BF scheme, the optimization problem (27) has  $N_h^2$ design variables and $2K + 2Q + 1$ slack variables, ${N_h} + 2M + 2K + 2Q$ LMI constraints of size 1, $K + Q + 1$ LMI constraints of size ${N_h}$, and 5 SOC constraints of size dimension  $N_h^2 + {N_h} + 1$. As a result, the computational complexity of the proposed robust BF scheme is $\begin{array}{l}
{\cal O}\left(\!\! {\sqrt {2M + \left( {{N_h} + 3} \right)K + \left( {{N_h} + 3} \right)Q + 2{N_h}}  \cdot n \cdot \left[ {\left( {{N_r} + 3M + 2} \right) \cdot } \right.}\!\! \right.\!\!\\
\left. {\left. \!\!{\left({1 + n} \right)\!\! +\!\! \left( {K\! + \!Q \!+ \!1} \right)\!\!\left( {{N_h} + n} \right)N_h^2 \!\!+\!\! \left( {K + Q} \right)\!\!\left( {N_h^2 + {N_h} + 2} \right)\!\! + {n^2}} \right]} \right)\!\!,
\end{array}$ where $n = {{\cal O}}\left( {N_h^2 + 2K + 2Q + 1} \right)$.

\vspace{-1em}
\section{Numerical Results}
\vspace{-0.3em}
In this section, numerical simulations are carried out to illustrate the performance of the proposed robust BF scheme. We consider a CSAN with  $Q = 2$ PUs, $M = 2$  SUs and $K = 3$  Eves. The achievable rate threshold of each SU is  ${\gamma _{th}} = 15\;{\rm{dB}}$ and the interference power threshold is ${{\rm{I}}_{th}} =  - 20\;{\rm{dB}}$. The  auto-correlation matrices of CSI error for Eves and PUs are  ${{\bf{E}}_{h,k}} = {\varepsilon _h}{{\bf{I}}_{{N_h}}}$ and ${{\bf{E}}_{g,q}} = {\varepsilon _g}{{\bf{I}}_{{N_h}}}$  with ${\varepsilon _h} > 0$  and ${\varepsilon _g} > 0$, respectively. According to ITU [\cite {19}], $\varphi _a^{{\rm{3dB}}}$ and $\varphi _e^{{\rm{3dB}}}$  can be set, respectively, as $70^\circ $ and $15^\circ $, and $SLL$ can be set as 20 dB. The outage probabilities for both of outage constraints are set as ${P_{out,1}} = {P_{out,2}} = 0.1$. The benchmark (perfect CSI), SDR combined with randomlization [\cite {5447068-17}]  and the non-robust BF scheme are compared with our proposed scheme. 

\begin{figure*} [htpb]
\vspace{-0.8cm}  
\setlength{\abovecaptionskip}{-0.12cm}   
\setlength{\belowcaptionskip}{-0.6cm}   
\begin{minipage}[t]{0.5\linewidth} 
\centering \includegraphics[width=9cm,height=7cm]{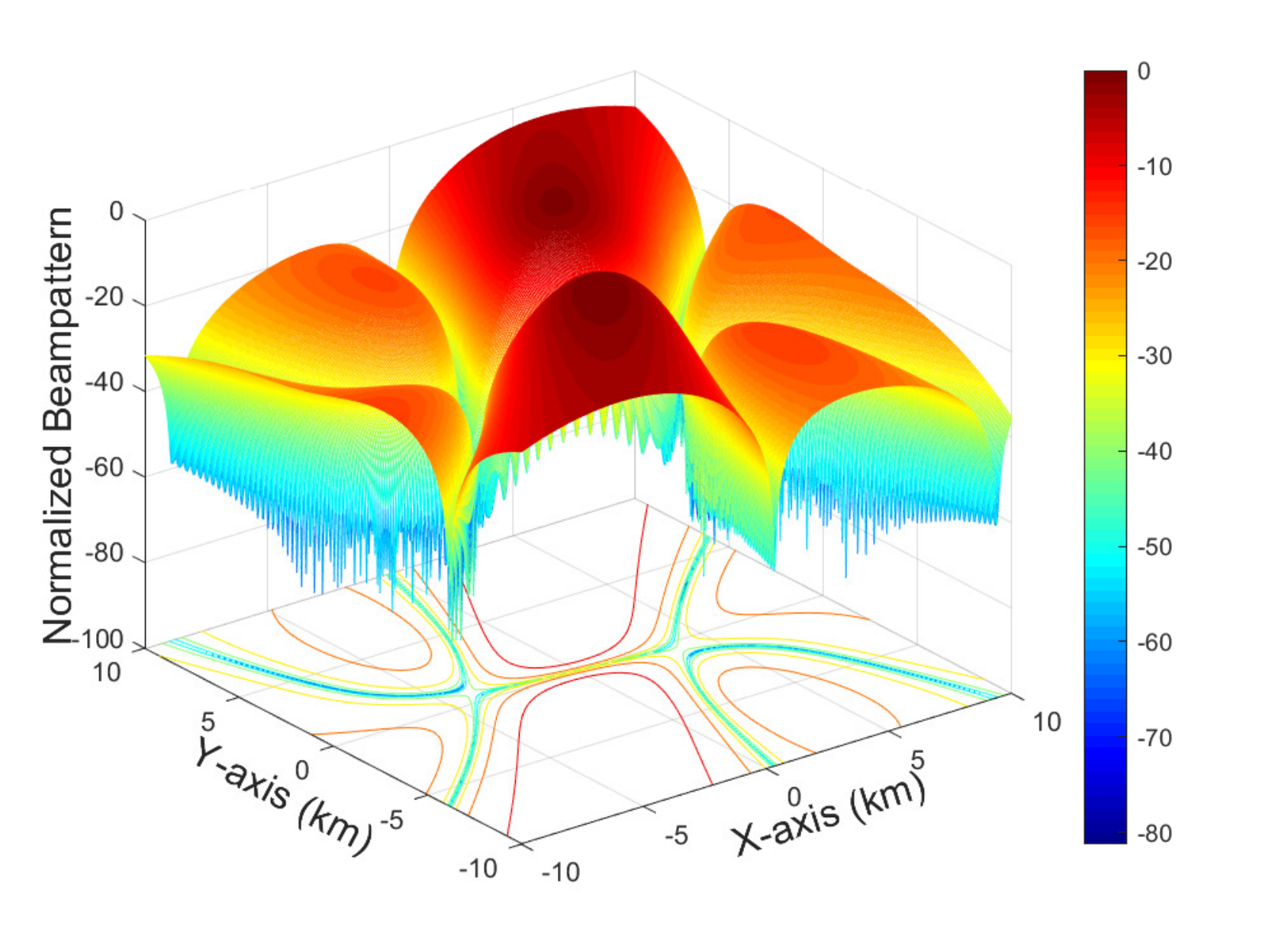} \caption{ 3D Beampattern of  ${\mathbf{w}}$} \label{fig:side:a} 
\setlength{\belowcaptionskip}{-1cm}
\end{minipage} 
\begin{minipage}[t]{0.5\linewidth} 
\centering \includegraphics[width=9cm,height=7cm]{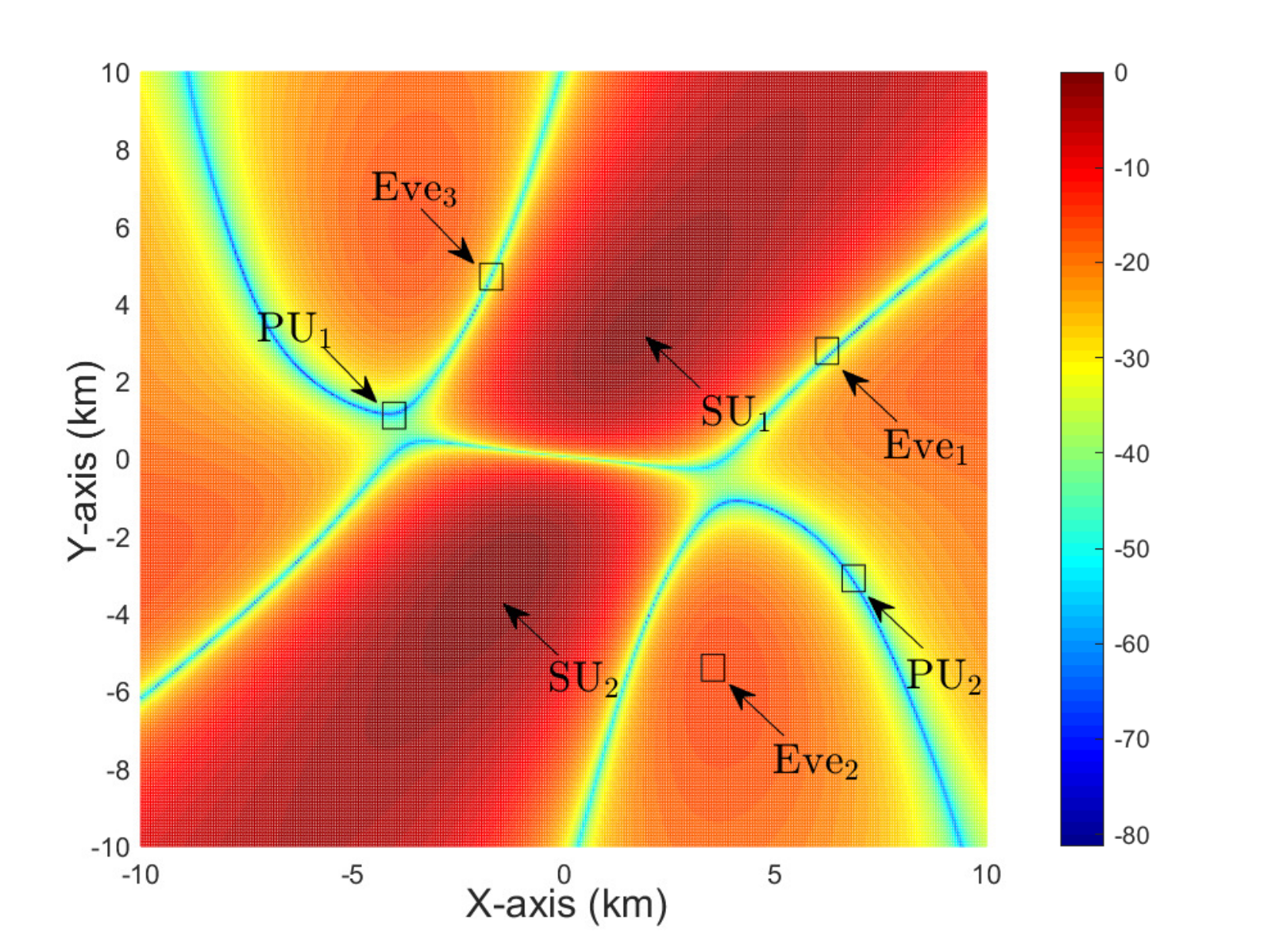} \caption{Beampattern of ${\mathbf{w}}$  from vertical view} \label{fig:side:b} 
\setlength{\belowcaptionskip}{-1cm}
\end{minipage}  
\end{figure*}

Fig. 1 and Fig. 2 illustrate the normalized beampattern of the beamforming weight vector ${\bf{w}}$. We can observe that the beampattern generates two wavepeaks at the locations of SUs. Furthermore, several nulls are generated at Eves and PUs with depths of -60 dB to -40 dB. Although a secondary lobe points to  ${\rm{Ev}}{{\rm{e}}_2}$, it still creates a level of suppression at -15 dB. These results demonstrate that the proposed robust BF scheme is capable of satisfying the SNR requirements of all intended users and degrading the signal strength of unintended users.

\begin{figure}[htbp]
\vspace{-0.45cm}  

\centering
  \includegraphics[width=12cm,height=9cm]{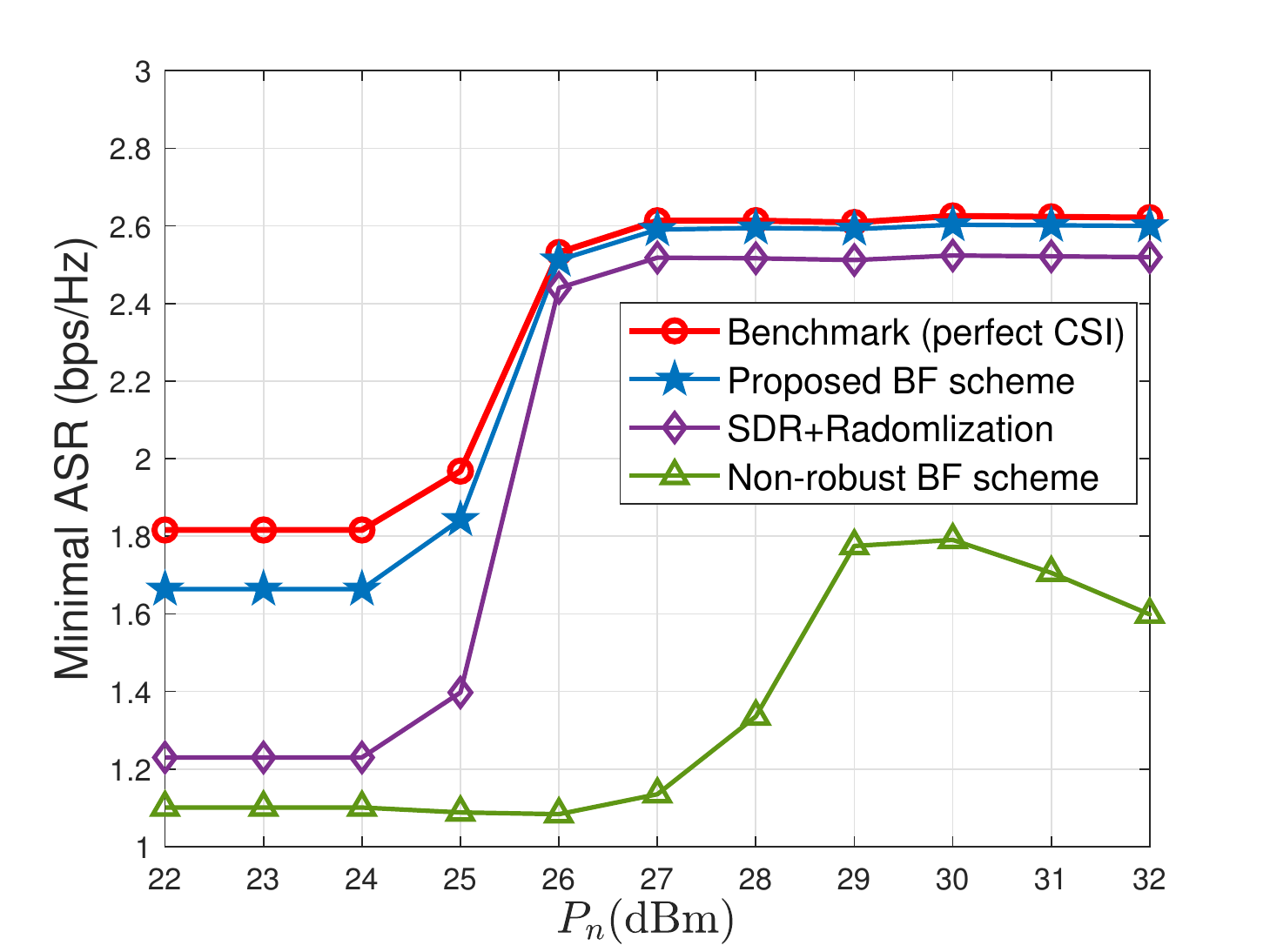}
  \caption{{Minimal ASR versus per-antenna power budget  ${P_n}$} \label{fig:side:a}}\label{FigOPPRPS}
\end{figure}


To further investigate the minimal ASR performance based on multicast communication, we investigate the minimal ASR versus per-antenna power with the robust BF, SDR-randomlization and non-robust BF schemes. As seen from Fig. 3, the proposed robust BF scheme significantly outperforms the SDR-randomlization and non-robust BF schemes, and the gap between the robust BF scheme and the benchmark scheme is gradually decreasing with increasing per-antenna power budget, which verifies the robustness of the proposed scheme. Moreover, one can observe that the minimal ASR of the non-robust scheme is decreased when the high per-antenna power is increased to a certain high value.
\begin{figure}[htbp]
\vspace{-0.45cm}  
\setlength{\abovecaptionskip}{-4cm}   
\setlength{\belowcaptionskip}{-0.4cm}   
\centering
  \includegraphics[width=12cm,height=9cm]{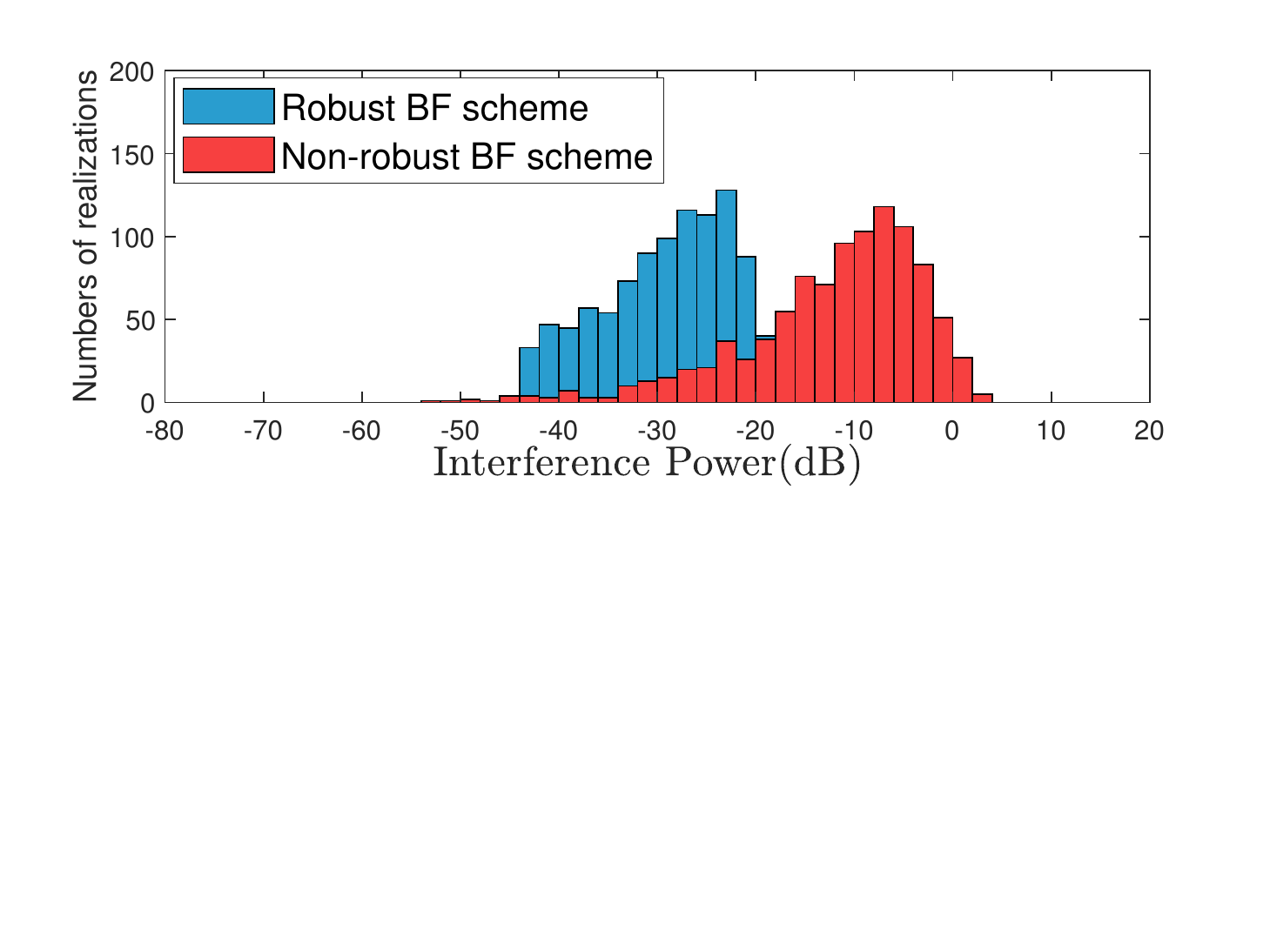}
  \caption{{Distribution histogram of interference power}}\label{FigOPPRPS}
\end{figure}

In Fig. 4, we plot the distribution histogram of interference power with respect to different BF schemes, after 1000 channel realizations. As expected, about $90\% $ of the interference power of the robust BF scheme is below the interference power threshold, i.e.  ${{\rm I}_{th}} =  - 20\;{\rm{dB}}$, while only $30\% $ of the interference power of the non-robust BF scheme satisfies the interference constraint, which confirms the robustness to channel error of our proposed BF scheme.

\vspace{-1em}
\section{Conclusion}
\vspace{-0.3em}
In this paper, we have presented a robust BF scheme to guarantee PLS in the considered CSAN. By employing only imperfect CSIs of Eves and PUs available at HAP, and adopting multicast transmission  at both satellite and HAP, we formulated a maximization problem for the minimal ASR of multiuser system. To handle this non-convex optimization problem, we applied an SDP method with a penalty function to convert the original problem into a convex one, which was then solved in an iterative manner. Simulation results show that the proposed robust BF scheme outperforms the benchmarks in terms of the suppression of unintended secondary users as well as the minimal ASR of the satellite-aerial network.




\appendices

\ifCLASSOPTIONcaptionsoff
  \newpage
\fi



%

\footnotesize
\bibliographystyle{IEEEtran} 
\bibliography{IEEEabrv,CL2021_ref} 

\end{document}